# Scaling of the Local Dynamics and the Intermolecular Potential

C.M. Roland[1], J.L. Feldman[1,2], and R. Casalini[1,3]

[1]*Naval Research Laboratory, Washington, DC 20375-5342*

[2]*George Mason University, School of Computational Sciences, Fairfax, VA 22030*

[3]*George Mason University, Chemistry Department, Fairfax, VA 22030*

## Abstract

The experimental fact that relaxation times, τ, of supercooled liquids and polymers are uniquely defined by the quantity $TV^{\gamma}$, where T is temperature, V specific volume, and γ a material constant, leads to a number of interpretations and predictions concerning the dynamics of vitrification. Herein we examine means to determine the scaling exponent γ apart from the usual superpositioning of relaxation data. If the intermolecular potential can be approximated by an inverse power law, as implied by the $TV^{\gamma}$ scaling, various equations are derived relating γ to the Grüneisen parameter and to a common expression for the pressure derivative of the glass temperature. In addition, without assumptions, γ can be obtained directly from pressure-volume-temperature data. These methods for determining γ from molecular or thermodynamic properties are useful because they enable the P- and V-dependences of τ to be obtained, and thereby various analyses of the dynamics to be explored, without the need to carry out relaxation measurements beyond ambient pressure.

## 1. Introduction

As demonstrated for more than 40 molecular and polymeric glass-forming materials [1,2,3,4,5,6], the local relaxation times (structural or segmental) measured under various thermodynamic conditions (e.g., different temperatures and pressures) in the equilibrium state (T > $T_g$(P)) superpose when plotted as a function of temperature times the specific volume, the latter raised to a material constant; that is,

$$\tau = \Im_1(TV^{\gamma}) \tag{1}$$

where $\Im_1$ represents some unknown function, which can be related to the excess configurational entropy [7]. Empirically, the exponent γ has been found to vary in the range 0.13 to 8.5 [8]. This power-law scaling is capable of describing relaxation times over the entire supercooled regime, from $T_g$ through temperatures beyond the dynamic crossover. Other forms for the scaling have been proposed (e.g. a linear scaling on density [9]), but these are valid only over a limited range [2,8]. Eq.(1) is an empirical fact without underlying assumptions. Of course, it would be useful to develop a theoretical basis for this dependence [7], in order to further understand the complex dynamics associated with vitrification. The first demonstration of eq.(1), for o-terphenyl (OTP) using γ = 4 [10,11,12], was motivated by the fact that the effective intermolecular potential of OTP is well-described by the Lennard-Jones (LJ) equation [13]

$$U(r) = Ar^{-12} - Br^{-6} \tag{2}$$

where A and B are constants and *r* is the intermolecular distance. This involves the approximation that a group of atoms comprising a molecule can be treated as a single particle. The exponent of 6 in the second term of eq.(2) is generally valid for van der Waals attractive interactions; however, the repulsive exponent varies among materials [14,15]. For dense liquids a common simplification is to include only the repulsive term, since for local properties the attractions primarily exert only a mean-field, density-dependent pressure. Such a minimum description arises from the fact that in a dense liquid the attractive forces from the many neighbors of a given molecule essentially cancel [16]; this simplification is also consistent with the fact that in liquids, the static structure factor at intermediate and large wave vectors is sensitive only to the repulsive part of the potential [17,18]. Dropping the attractive term and generalizing the LJ potential leads to an inverse-power-law (IPL) [19,20,21,22]

$$U(r) = Ar^{-3\gamma} \tag{3}$$

From such a (purely repulsive) potential, thermodynamic properties such as the total energy, volume, and entropy can be expressed functions of $r^{3\gamma}$ or $V^{\gamma}$ [20]. Thus, the IPL serves as an underpinning for the scaling of eq.(1), with γ employed as a material-constant parameter used to obtain superpositioning of experimental τ(T,V) data [1].

The assumptions of spherical symmetry and an absence of strong attractive forces, implicit in the IPL potential, are not obviously justified. Certainly the interactions are not spherically symmetric in polymers given the plethora of covalent intramolecular bonds or in associated liquids with strongly directional bonds. One justification for interpreting the scaling in terms of the IPL is that these strongly directional interactions are relatively insensitive to pressure (as shown for polyoxybutylene [23]). This means that they are essentially constant with respect to the $TV^{\gamma}$ scaling variable. This reasoning may also justify the very low values found for $\gamma$ for some H-bonded liquids [8], since only a fraction of the potential actually senses volume changes.

From this $TV^{\gamma}$ scaling a number of features of the glass-forming dynamics can be described or predicted: (i) the Boyer-Spencer-Bondi rule [3] that the product of the thermal expansion coefficient, $\alpha_P$ ($= V^{-1}\left.\frac{dV}{dT}\right|_P$), and the glass transition temperature is approximately a universal constant for polymers (i.e., $\alpha_P T_g \sim 0.18$) [24]; (ii) the constancy (pressure independence) for a given material of the relaxation time at the dynamic crossover [25]; (iii) the decrease in fragility with pressure for all non-associated organic glass-forming liquids and polymers [25]; and (iV) the nature of the T-, P-, and V-dependences of the normal mode of polymers [23]. Despite these successful developments from the dynamic scaling, it remains of interest to explore the origin of the scaling exponent. The parameter γ cannot be deduced from consideration of the molecular structure (of course, the same is true for other dynamic, as well as thermodynamic, properties). Thus, how can γ be obtained other than by empirical superposition of relaxation times? We describe herein routes to the determination of γ from molecular properties or thermodynamic (equation of state, EOS) data.

## 2. Results

### 2.1 Grüneisen parameter

Notwithstanding the limitations of the IPL description of the intermolecular potential described above, the IPL potentially offers a connection to the scaling exponent γ. The Grüneisen parameter, which is a measure of the anharmonicity of lattice vibrations, is defined formally in terms of the volume dependence of the phonon frequency, ω [26]

$$\gamma_G = -\frac{d\ln\omega}{d\ln V} = \frac{rd\omega}{3\omega dr} \tag{4}$$

This mode Grüneisen parameter [27] can be related to the intermolecular potential through the force constant, $K \equiv \frac{d^2U}{dr^2}$, which in the harmonic approximation is proportional to $\omega^2$. Eq.(3) then gives [28,29]

$$\gamma_G = \frac{1}{2}\gamma + \frac{1}{3} \tag{5}$$

Thus, through the parameter $\gamma_G$ the scaling exponent can be connected to the steepness of the intermolecular potential. In the vicinity of $T_g$, $\gamma_G$ shows only small variation with temperature [30,31], falling in the range $1 \le \gamma_G \le 4$. From eq.(5) this corresponds to $\gamma$ in the range from 2 to 9, consistent with experimental values of the scaling exponent [8]. Since $\gamma_G$ is defined in terms of phonon frequencies, the relevant value is for the solid (glassy) state, even though the scaling exponent is determined from measurements on the equilibrium liquid.

There is an alternative method of defining the Grüneisen parameter. From eq.(4) $\gamma_G$ can be expressed in terms of thermodynamic properties [30]

$$\gamma_G = \frac{V\alpha_P}{C_V\kappa_T} \tag{6}$$

in which $C_V$ is the isochoric heat capacity and $\kappa_T$ is the isothermal compressibility. $\gamma_G$ calculated using eq.(6) were reported in ref. [7]. These are reproduced in Table 1, along with the scaling exponent; the agreement is good considering the approximate nature of the analysis.

Substituting for $\alpha_P$, $C_V$, and $\kappa_T$ and using the Maxwell relations

$$\gamma_G = (V/T)\left.\frac{dS}{dV}\right|_T \Big/ \left.\frac{dS}{dT}\right|_V \tag{7}$$

As stated above, when eq.(3) is valid the thermodynamic properties, the entropy can be expressed as a function of $V^\gamma$; thus [20]

$$S(T,V) - S_{ideal} = \Im_2\left(TV^\gamma\right) \tag{8}$$

where $S_{ideal}$ is the ideal gas entropy and $\Im_2$ is an unknown function. Taking the derivatives in eq.(7) and noting that

$$dS_{ideal} = \frac{C_{V,ideal}}{T} dT + \frac{k}{V} dV \tag{9}$$

where k is Boltzmann constant and $C_{V,ideal} = 3k/2$, we obtain

$$\gamma_G = \frac{T\gamma V^{\gamma} \Im_2' + k}{TV^{\gamma} \Im_2' + 3k/2} \tag{10}$$

Since k is small compared to $TV^{\gamma}\Im'$, we find that $\gamma_G \sim \gamma$. Explicit solution of eq.(7) using a functional form for eq.(1) recently derived from an entropy model [32] gives the exact result $\gamma_G = \gamma$ [7].

## 2.2 PVT measurements

From eq.(1) it is straight forward to show that [2]

$$\gamma = -\frac{1}{T\alpha_{\tau}(T)} \tag{11}$$

in which $\alpha_\tau$ is the isochronic thermal expansion coefficient, $\alpha_{\tau} = V^{-1} \left.\frac{dV}{dT}\right|_{\tau}$. This equation has been verified for 19 different glass-formers, with $\alpha_\tau$ usually determined from relaxation measurements together with the EOS for the liquid [2]. Since at $T_g$ the relaxation time is constant (as seen from relaxation measurements carried out at elevated pressures [8]), eq.(11) becomes

$$\gamma = \left( T_g V_g^{-1} \frac{dV_g}{dT_g} \right)^{-1} \tag{12}$$

where $V_g$ refers to the specific volume at the glass transition. If PVT data are available to sufficiently low temperatures (< $T_g$), $\gamma$ can be obtained without measuring $\tau$.

We illustrate this in Figure 1 with PVT data for diglycidylether of bisphenol A (DGEBA) having a degree of polymerization equal to 5 [33]. The glass transition temperature at each pressure is determined from the intersection of linear fits to the liquid and glass data, yielding $\alpha_\tau(T_g)$. Table 2 shows a comparison of the $\gamma$ calculated from PVT measurements using eq.(12) compared to the value obtained by superposition of relaxation times.

## 2.3 Pressure dependence of $T_g$

The empirical relation of Simon [34] for the pressure dependence of the melting temperature has the form

$$P - P_t = a\left[\left(\frac{T}{T_t}\right)^c - 1\right] \tag{13}$$

where T and P correspond to the melting condition (solid and liquid in equilibrium), $P_t$ and $T_t$ refer to the triple point of the liquid (solid, liquid, and gaseous phase in equilibrium) and, a and c are material constants. As shown by Hoover and Ross [20], the Simon equation follows directly from the properties of the IPL, with $c = 1 + \gamma^{-1}$. Eq.(3) also gives for the EOS [20]

$$\frac{PV}{T} = \Im_3(TV^\gamma) \tag{14}$$

where $\Im_3$ is an unknown function. However, this expression is inaccurate due to the neglect of the longer-ranged attractions. As a first approximation eq.(14) can be modified to include a constant background pressure, $P_0$

$$\frac{(P + P_0)V}{T} = \Im_3(TV^\gamma) \tag{15}$$

We choose this form herein for the EOS because it leads to an equation for the pressure dependence of $T_g$ that has the form of the Simon equation. To show this, note that the glass transition corresponds to a fixed value of τ; thus, $V(T_g) = \text{constant} / T_g^{1/\gamma}$, which substituted into eq.(15) yields

$$T_g(P) = T_g(0)\,(1 + P_0^{-1}\,P)^{\gamma/(\gamma+1)} \tag{16}$$

Eq.(16), used empirically to describe the pressure dependence of the glass transition [8,35,36], is identical in form to the Simon equation.

In Figures 2-4 experimental PVT data are plotted for three glass-formers [23,37,38], chosen because they span a range of γ. As can be seen, eq.(15) provides a reasonable approximation for the EOS, using the γ determined from superpositioning of τ(T,P) data. $P_0$ is the only adjustable parameter. From eq.(16) the pressure coefficient of $T_g$ in the limit of low pressure is given by

$$\lim_{P \to 0} \frac{dT_g}{dP} = \frac{\gamma}{(\gamma + 1)} \frac{T_g(0)}{P_0} \tag{17}$$

In Table 3 we compare the pressure coefficient calculated from eq.(17) to experimental values; the results are in good accord. At higher pressures, well beyond the range of the PVT measurements (which are limited to < 200 MPa), $T_g$ decreases more strongly than predicted by eq.(16).

### 3. Concluding remarks

The experimental finding that structural relaxation times for glass-forming liquids and polymers can be superimposed when plotted versus of $TV^{\gamma}$ suggests that for local dynamics the intermolecular potential function can be approximated by an IPL. If correct, this leads to the expectation of a relationship between the scaling exponent $\gamma$ and other properties. One example is the connection between $\gamma$ and the Grüneisen parameter. The quantity $\gamma_G$ describes vibrational modes for the solid state, and eqs. (4) and (6) are strictly valid only for solids. However, all liquids are solid-like at short times [39]; thus, the concept of a Grüneisen parameter may be appropriate even for the liquid state. Formally the Grüneisen parameter characterizes the anharmonicity of the molecular vibrations, and there have been previous efforts to relate phonon anharmonicity to structural relaxation [40,41,42]

We show that the exponent yielding superpositioning of $\tau(T,V)$ data is close to the value of the Grüneisen parameter, the exact relationship being model-dependent. An IPL intermolecular potential yields eq.(5), which in turn gives reasonable values for $\gamma$ (Table 1). However, there are obvious limitations of this analysis: The relationship between $\gamma_G$ and $\gamma$ is exact for an IPL potential but the latter only approximates the forces between real, non-spherical molecules. Moreover, it has been found that even when the IPL provides an accurate description, the exponent ($3\gamma$ in eq.(3)) can decrease with increasing temperature [43]. The $\tau(TV^{\gamma})$ scaling is predicated on $\gamma$ being constant. Describing the supercooled dynamics with an entropy model [7], a some what different relation is obtained, $\gamma = \gamma_G$. This underestimates the scaling exponent in comparison to eq.(5), although reliable values for $\gamma_G$ are scarse.

Following from the IPL potential, the Simon equation for the pressure derivative of the melting point is obtained, with an analogous equation for the glass transition temperature (eq.(16)) used empirically [8,35,36]. We show herein that the latter can be derived provided that a specific form for the EOS (eq.(15)) is assumed. To the extent this EOS gives an accurate description of experimental PVT data, values for $dT_g/dP$ can be used to estimate $\gamma$ from eq.(17).

Notwithstanding its theoretical basis, the scaling exponent can be determined directly from PVT data via eq.(12); that is, $\gamma$ is obtained without recourse to superpositioning of experimental τ(T,V). Accordingly, relaxation times obtained at ambient pressure can be extended to elevated pressures (or to isochoric conditions) by utilizing the scaling relationship. The only other requirement is knowledge of the EOS, which is deduced from the same PVT data yielding the scaling exponent.

**Acknowledgements**

This work was supported by the Office of Naval Research.

**References**

[1] R. Casalini, C.M. Roland, Phys. Rev. E 69 (2004) 062501.

[2] C.M. Roland, R. Casalini, J. Non-Cryst. Solids 251 (2005) 2581.

[3] R. Casalini, C.M. Roland, Coll. Polym. Sci. 283 (2004) 107.

[4] C. Dreyfus, A. LeGrand, J. Gapinski, W. Steffen, A. Patkowski, Eur. Phys. J. 42 (2004) 309.

[5] C. Alba-Simionesco, A. Cailliaux, A. Alegria, G. Tarjus, Europhys. Lett. 68 (2004) 58.

[6] C.M.Roland, R.Casalini, J. Chem. Phys. **121**, 11504 (2004)

[7] R. Casalini, U. Mohanty, and C.M. Roland, J. Chem. Phys., submitted.

[8] C.M. Roland, S. Hensel-Bielowka, M. Paluch, R. Casalini, Rep. Prog. Phys. 68 (2005) 1405.

[9] G. Tarjus, S. Mossa and C. Alba-Simionesco, J. Chem. Phys. 121 (2004) 11505.

[10] A. Tölle, Rep. Prog. Phys. 64 (2001) 1473.

[11] C. Dreyfus, A. Aouadi, J. Gapinski, M. Matos-Lopes, W. Steffen, A. Patkowski, R.M. Pick, Phys. Rev. E 68 (2003) 011204.

[12] G. Tarjus, D.Kivelson, S.Mossa, C.Alba-Simionesco, J.Chem.Phys. 120 (2004) 6135.

[13] L.J. Lewis, G.Wahnström, Phys. Rev. E 50 (1994) 3865.

[14] E.A. Moelwyn-Hughes, Physical Chemistry, 2nd edition, Pergamon Press, New York (1961).

[15] V.Y. Bardik and V.M. Sysoev, Low Temp. Phys. 24 (1998) 601.

[16] B. Widom, Physica A 263 (1999) 500.

[17] J.D. Weeks, D. Chandler, H.C. Andersen, J. Chem. Phys. 54 (1971) 5237.

[18] A. Tölle, H. Schober, J. Wuttke, O.G. Randl, F. Fujara, Phys. Rev. Lett. 80 (1998) 2374.

[19] H.C. Longuet-Higgins, B. Widom, Mol. Phys. 8 (1964) 549.

[20] W.H. Hoover, M. Ross, Contemp. Phys. 12 (1971) 339.

[21] R.J. Speedy, J. Phys. Cond. Mat. 15 (2003) S1243.

[22] M.S. Shell, P.G. Debenedetti, E.La Nave, F. Sciortino, J. Chem. Phys. 118 (2003) 8821.

[23] R. Casalini, C.M. Roland, Macromolecules 38 (2005) 1779.

[24] D.W. Van Krevelen, Properties of Polymers, Elsevier, NY (1990).

[25] R. Casalini and C.M. Roland, Phys. Rev. B 71 (2005) 014210.

[26] E. Grüneisen, Ann. Phys. (Leipzig) 39 (1912) 257.

[27] The assumption of a single mode is an approximation; more generally the value of $\gamma_G$ is mode-dependent.

[28] K.A. Moelwyn-Hughes, J. Phys. Coll. Chem. 55 (1951) 1246.

[29] J.R. Hook, H.E. Hall, Solid State Physics, 2nd Edition, J. Wiley & Sons, W. Sussex, United Kingdom, 1991.

[30] G. Hartwig, *Polymer Properties at Room and Cryogenic Temperatures* (Plenum Press, New York, 1994), chapter 4.

[31] J.G. Curro, J.Chem. Phys. 58 (1973) 374.

[32] I. Avramov, J. Non-Cryst. Solids 262 (2000) 258.

[33] M. Paluch, C.M. Roland, J. Gapinski, and A. Patkowski, J. Chem. Phys. 118 (2003) 3177.

[34] F.E. Simon, G. Glatze, Z. Anorg. Allgem. Chem. 178 (1920) 309.

[35] S. P. Andersson, O. Andersson, Macromolecules 31 (1999) 2999.

[36] A. Drozd-Rzoska, Phys. Rev. E 72 (2005) 041505.

[37] C.M. Roland, R. Casalini, P. Santangelo, M. Sekula, J .Ziolo and M. Paluch, Macromolecules 36 (2003) 4954.

[38] C.M. Roland and R. Casalini, J. Chem. Phys. 122 (2005) 134505.

[39] T. Keyes, J. Phys. Chem. A 101 (1997) 2921.

[40] C.A. Science 267 31 (1995) 1924.

[41] J.C. Dyre, N.B. Olsen, T. Christensen, Phys. Rev. B 53 (1996).

[42] C.M. Roland and K.L. Ngai J. Chem. Phys.104 (1996) 2967.

[43] V.Y. Bardic, N.P. Malomuzh, V.M. Sysoev, J. Mol. Liq. 120 (2005) 27.

Table 1. Scaling Exponent from the IPL

| | $\gamma_G$ [7] | $\gamma$ | |
|---|---|---|---|
| | | eq.(5) | eq.(1) [8] |
| OTP | 1.2 | 1.7 | 4 |
| polyvinylacetate | 0.7 | 0.7 | 2.5 |
| polymethylmethacrylate | 0.7 | 0.7 | 1.25 |
| salol | 1.9 | 3.1 | 5.2 |
| propylene carbonate | 1.4 | 2.1 | 3.7 |

Table 2. Scaling Exponent from PVT

| | $\gamma$ | |
|---|---|---|
| | eq.(1) [8] | eq. (12) |
| KDE* | 4.5, 4.8 | 4.8 |
| DGEBA | 2.8, 3.6 | 4.0 |
| polymethylmethacrylate | 1.25 | 1.9 |

*cresolphthalein-dimethylether

Table 3. Pressure coefficient of $T_g$

| | | | | $\lim_{P\to 0}\frac{dT_g}{dP}$ (K/MPa) | |
|---|---|---|---|---|---|
| | $\gamma$ [8] | $T_g(0)$ (K) [8] | $P_0$ (MPa) | experimental [5] | eq.(17) |
| polyvinylethylene | 1.9 | 253 | 900 | 0.240 | 0.18 |
| polyoxybutylene | 2.65 | 199 | 850 | 0.155 | 0.17 |
| polychlorinated biphenyl | 8.5 | 277 | 650 | 0.310 | 0.38 |

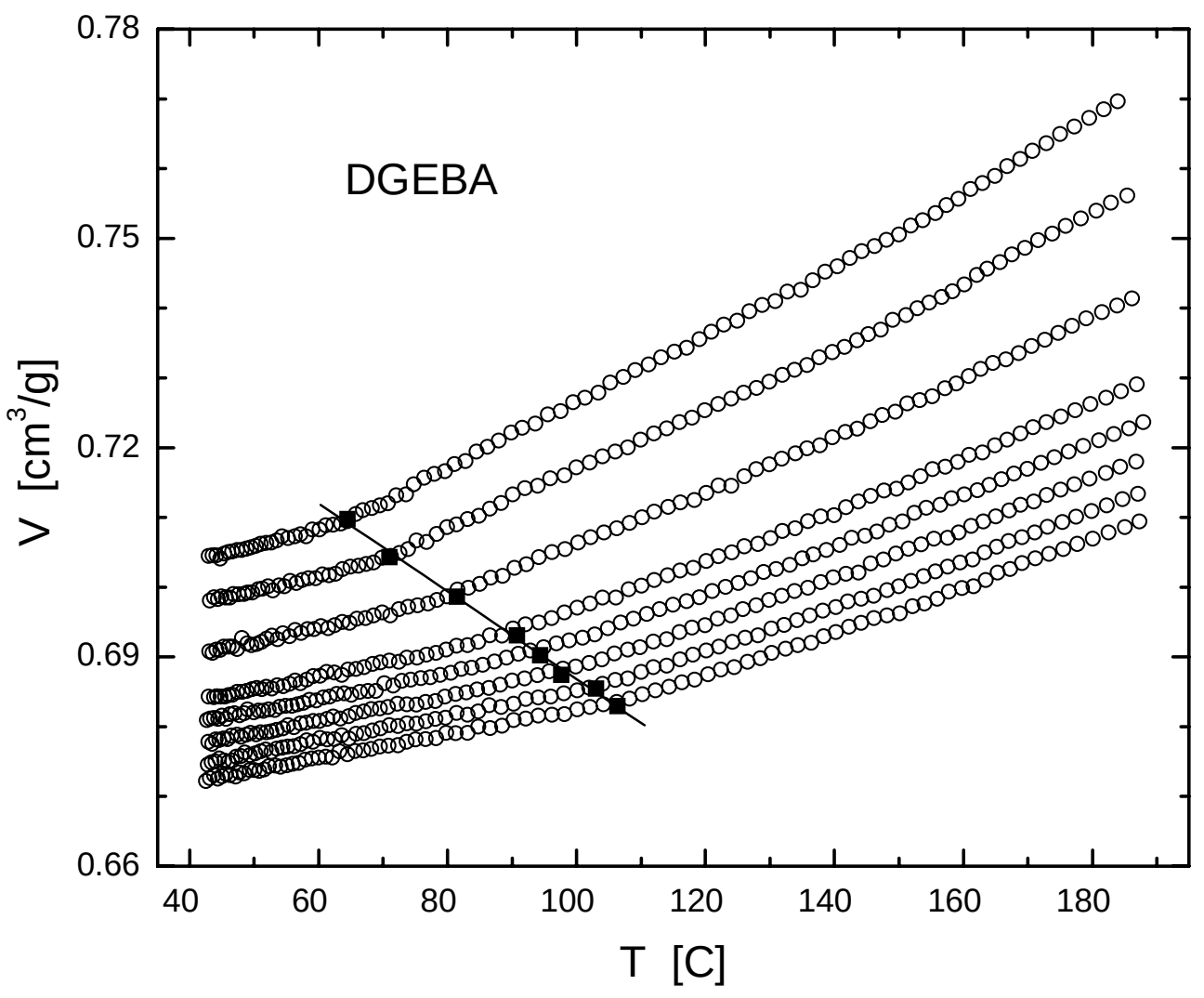

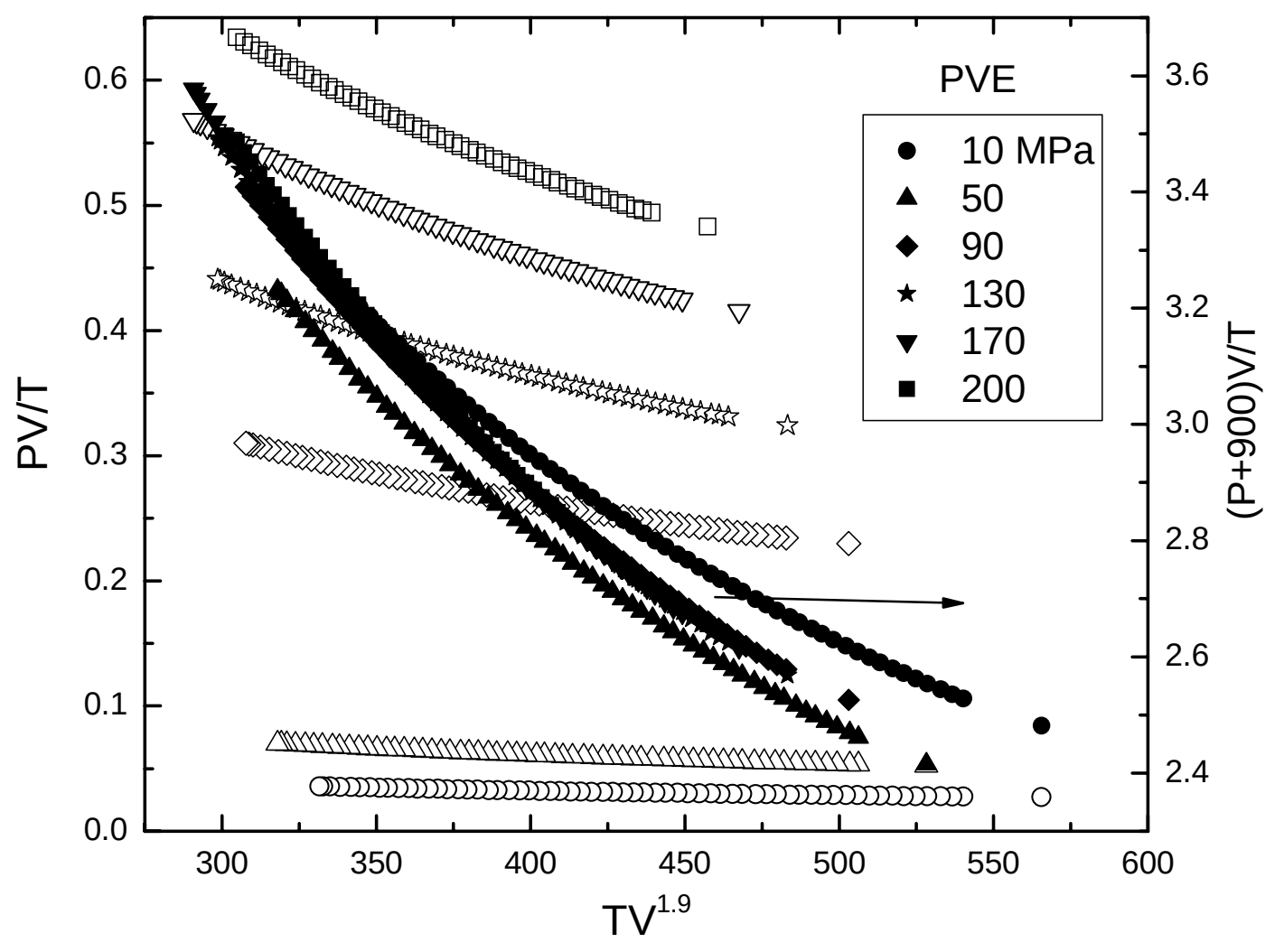

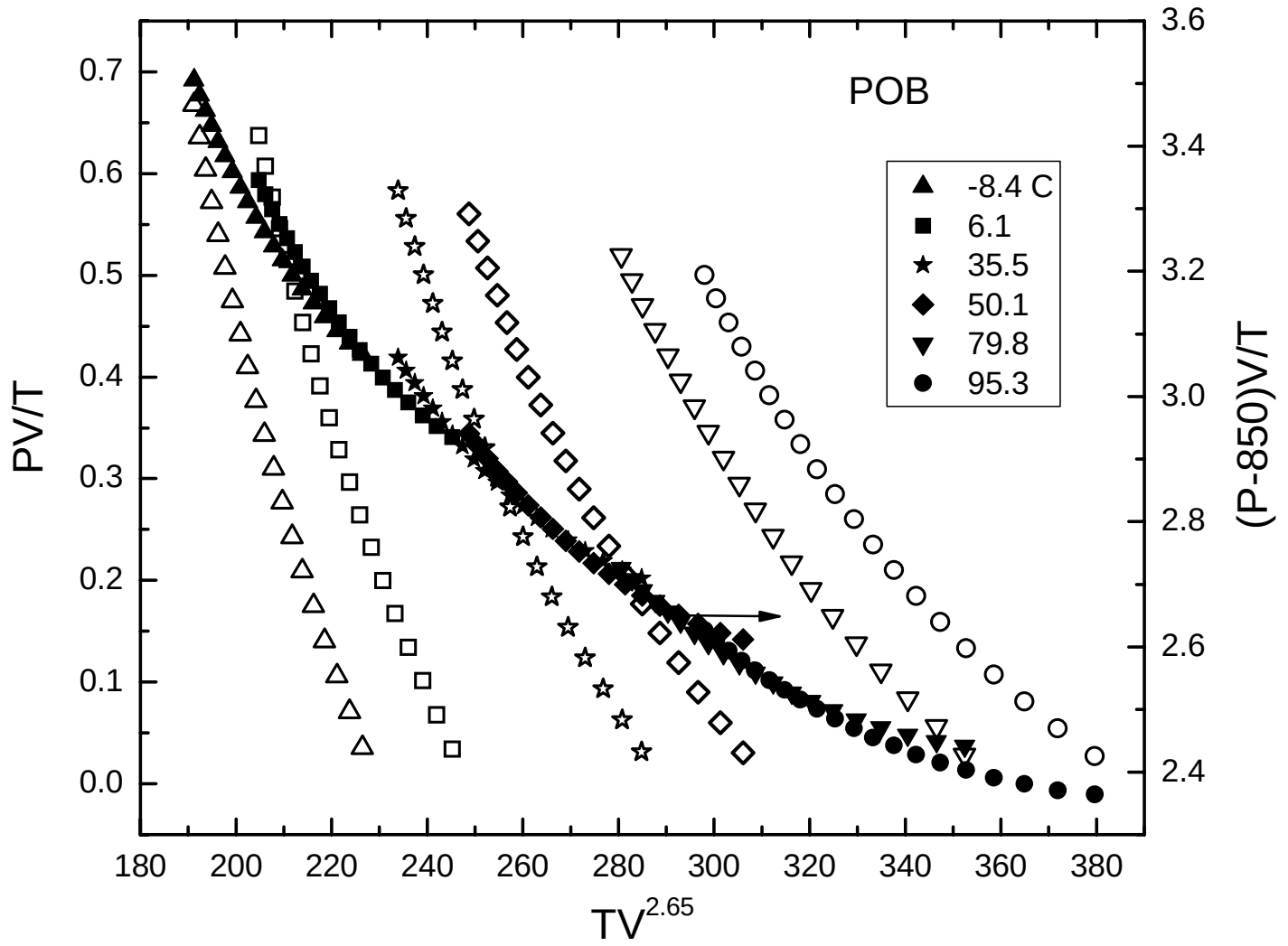

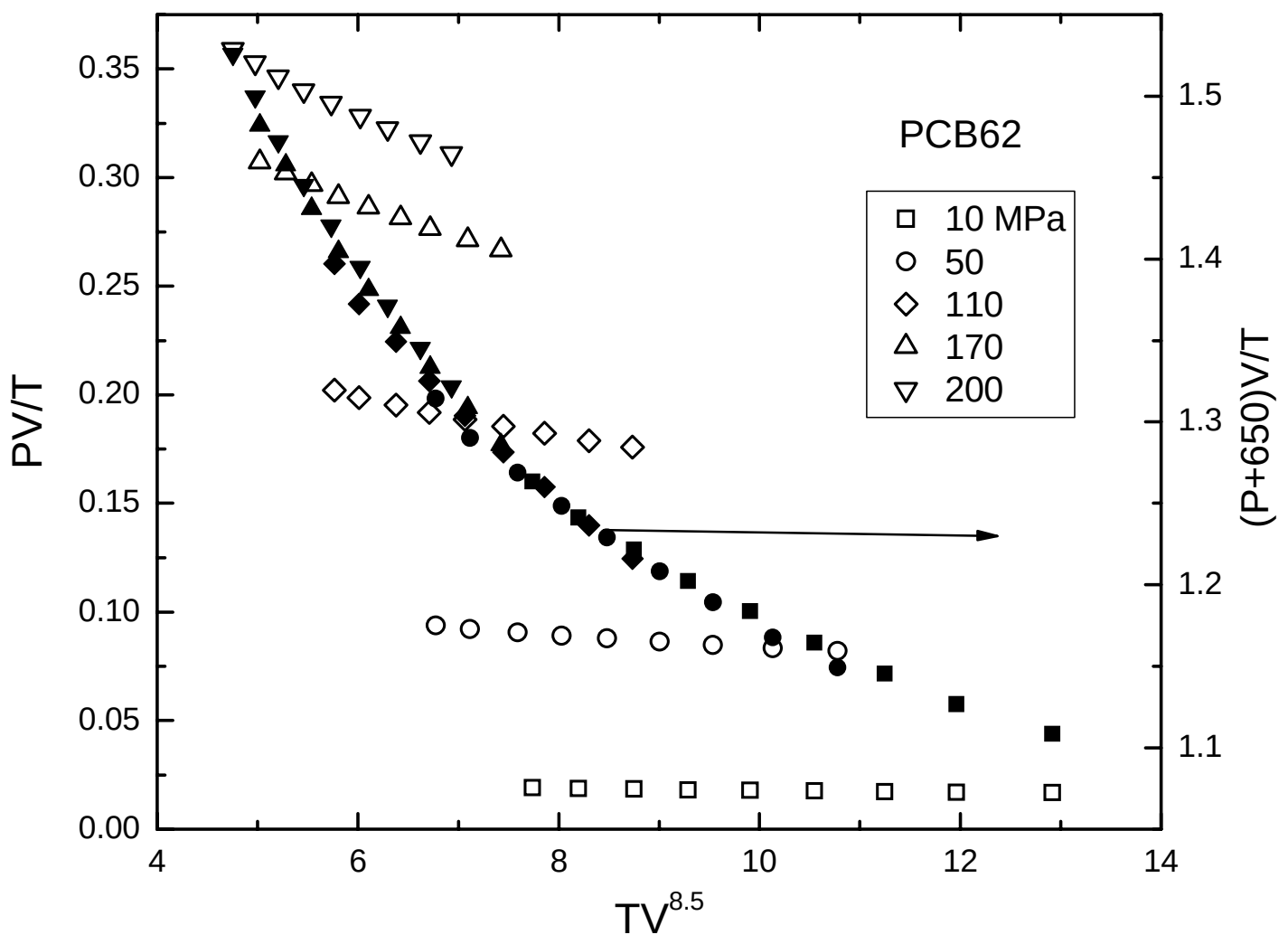

## Figure Captions

Figure 1. PVT data for DGEBA [33]. The solid symbols are $V_g$, the T-dependence of which yields $\alpha_\tau$ and via eq.(12) the scaling exponent $\gamma$.

Figure 2. EOS suggested by eq.(14) (hollow symbols) and eq.(15) or polyvinylethylene [37].

Figure 3. EOS suggested by eq.(14) (hollow symbols) and eq.(15) for polyoxybutylene [23].

Figure 4. EOS suggested by eq.(14) (hollow symbols) and eq.(15) for chlorinated biphenyl (62% by weight chlorine) [38].